\documentclass[aps,prl,twocolumn,showpacs,groupedaddress]{revtex4} 
\usepackage[american]{babel}
\usepackage{color,graphicx,amsmath,amssymb,bm}
\newcommand\be{\begin{equation}}
\newcommand\ee{\end{equation}}
\newcommand\bea{\begin{eqnarray}}
\newcommand\eea{\end{eqnarray}}

\newcommand\ket[1]{\left|#1\right\rangle}
\newcommand\bra[1]{\left\langle#1\right|}

\newcommand\Avg[1]{\langle#1\rangle}

\newcommand\ra{\rightarrow}

\newcommand\trm{\textrm}

\newcommand\id{{\rm 1} 
        \hspace{-1.1mm} {\rm I}
        \hspace{0.5mm}}

\newcommand\eq[1]{Eq.~(\ref{#1})}

\begin{document}

\title{Weak values of electron spin in a double quantum dot}     
  
\author{Alessandro Romito}
\affiliation{
  Department of Condensed Matter Physics, 
  The Weizmann Institute of Science,     
  Rehovot 76100, Israel}

\author{Yuval Gefen}
\affiliation{
  Department of Condensed Matter Physics, 
  The Weizmann Institute of Science,     
  Rehovot 76100, Israel}

\author{Yaroslav M. Blanter}
\affiliation{
 Kavli Institute of Nanoscience
Delft University of Technology
Lorentzweg 1,  2628 CJ Delft,
The Netherlands}
\affiliation{
Department of Condensed Matter Physics, 
  The Weizmann Institute of Science,     
  Rehovot 76100, Israel}

\date{\today}

\begin{abstract}
We propose a protocol for a controlled experiment to measure a weak value of the electron's spin in a solid state device. 
The weak value is obtained by a two step procedure -- weak measurement followed by a strong one (post-selection), 
where the outcome of the first measurement is kept provided a second post-selected outcome occurs. 
The set-up consists of a double quantum dot and a weakly coupled quantum point contact to be used as a detector. 
Anomalously large values of the spin of a two electron system are predicted, as well as negative values of 
the {\it total} spin. We also show how to incorporate the adverse effect of decoherence into this procedure.
\end{abstract}

\maketitle

{\it Introduction.}--- The measurement of any observable in quantum mechanics is a probabilistic process 
described by the projection postulate~\cite{Neumann1932}. 
Each eigenvalue of the observable happens to be an outcome of the measurement process with a given probability, 
and the original state of the system collapses into the corresponding eigenstate. 
An intriguing viewpoint of quantum mechanics, based on a two-vector 
formulation~\cite{Aharonov:1964aa}, stipulates that the measured  
value of an observable depends on both a ``past vector'',  $\ket{\chi_0}$,
 at which the system is prepared (pre-selection), and a ``future vector'',  
$\ket{\chi_f}$, where a given state is selected following the measurement (post-selection).
Within this framework a procedure that leads to a {\it weak value} (WV)~\cite{Aharonov:1988aa} 
involves a weak measurement (i.e. a measurement that disturbs the system weakly) 
whose outcome is kept provided a second postselected outcome occurs.
The protocol for the WV of $\hat{A}$,
${}_f\Avg{\hat{A}}_0^{(W)}=\Avg{\chi_f|\hat{A}|\chi_0}/\Avg{\chi_f|\chi_0}$~\cite{Aharonov:1988aa,Aharonov:1990aa}, involves three steps:
(i) preselection, (ii) weak measurement of $\hat{A}$, 
and then (iii) projective post-selection of an eigenstate of  observable  $\hat{B}$. 
WVs can be orders of magnitude larger than strong values~\cite{Aharonov:1988aa}, 
negative (where conventional strong values would be positive definite)~\cite{2002proc.}, 
or even complex. 
Such non-standard values open an intriguing window to some fundamental aspects of 
quantum measurement, including access to simultaneous measurement of 
non-commuting variables~\cite{non_commuting}; 
dephasing and phase recovery~\cite{Neder:2007aa}; correlation between measurements~\cite{correlations}; 
and even new horizons in metrology~\cite{Aharonov:1988aa}. 

While some aspects of WVs have been demonstrated in optics 
based setups~\cite{optics}, the arena of quantum solid state  
offers very rich physics to be studied  through a WV measurement, and the possibility 
to fine tune and control the system's parameters through electrostatic gates and an 
applied magnetic field. 
The three main challenges in such an undertaking are: (1) overcoming the adverse 
effects of  dephasing during the application of the protocol, (2) designing a 
tunable detector that will operate in  both the strong and weak measurement regimes,
 (3) design a protocol such that step (ii) and (iii) do not commute, 
 \be 
 [\hat{A},\hat{B}]\neq0 \, ,
 \label{uno}
 \ee
 notwithstanding the fact that both $\hat{A}$ and $\hat{B}$ 
address the charge degree of freedom.     

Here we propose a protocol for a controlled experiment to measure the 
weak value of  the electron's spin. Our protocol overcomes the above
 mentioned difficulties. We also show how to incorporate the adverse effect of 
decoherence into this procedure. 
The set-up consists of a double quantum dot, recently studied as a 
candidate for a quantum computer qubit~\cite{Petta:2005aa}, 
and a weakly coupled quantum point contact to be used as a detector. 

Within the protocol alluded to above, the interaction 
between the detector (D) and the system (S) is weak in the coupling parameter  $\lambda \ll1$. 
As a result of the weak measurement, the shift in the detector's coordinate,  $q$,
 is  $\propto \lambda$, while the modification of the state of the system is 
 $\propto \lambda^2$, hence $\ket{\chi_0}$  is unchanged to order  $\lambda$. 
In an ideal strong measurement there is a one-to-one correspondence between 
the observed value of the detector's coordinate,  $q_{\alpha}$, and the state of S,
 $\ket{\alpha}$. 
Within a weak measurement procedure the ranges of values of $q$  that correspond
 to two distinct states of S, $\ket{\alpha}$  and $\ket{\alpha'}$,  are described 
by two probability distribution functions, $P_{\alpha}(q)$  and $P_{\alpha'}(q)$  
respectively. These distributions strongly overlap. Hence the measurement of $q$ 
 provides only partial information on the state of S.

{\it The model.}--- Petta {\it et al.}~\cite{Petta:2005aa} studied experimentally 
a device that  can operate and be controlled over time scales up to tens of 
nanoseconds or more, preserving quantum coherence. 
This system (cf. Fig. 1(a)) consists of a gate confined semiconducting 
double quantum dot hosting two electrons. 
\begin{figure}[ht!]
\label{fig_1}
\begin{center}
\includegraphics[width=70mm]{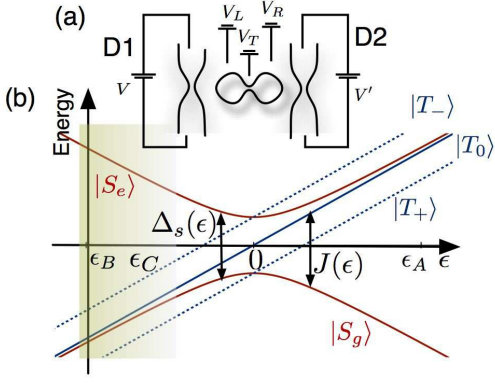}
\end{center}
\caption{(Color online) Schematics of the system and energy levels. 
(a) Scheme of a double dot with nearby quantum point contacts (QPCs) as charge sensors.
(b) Energy levels of lowest singlet (red) and triplet (blue) states vs. the detuning parameter $\epsilon$. In the $(0,2)$ charge configuration the antisymmetric nature of the electrons wave function implies a singlet ground state. The  states  $\ket{T_{\pm}}$ (blue dashed lines), with angular momenta $\pm \hbar$  in the direction of the applied magnetic field, are split by the Zeeman energy ($\Delta=g \mu_B B \approx 2.5 \mu eV$  in Ref. 4). The range of $\epsilon$ in which the effect of nuclear interaction is relevant is highlighted by the shadowed part.
Spin-to-charge conversion. If the variation $\epsilon_B \ra \epsilon_A$  is fast on the time scale of the nuclear field coupling, $\Avg{T_0(1,1)|H_N|S(1,1)}$, (''fast adiabatic''), $\ket{S(1,1)}$ is mapped into the ground state, $\ket{S(0,2)}$. For the ''slow adiabatic'' limit the ground state of the total Hamiltonian, $\ket{\uparrow \downarrow}$, will be mapped into $\ket{S(0,2)}$. By measuring the charge of the final state one is effectively measuring either the singlet, triplet or the $\ket{\uparrow \downarrow}$, $\ket{\downarrow \uparrow}$ component of the initial state at  $\epsilon=\epsilon_B$, depending on the specific time sequence of  $\epsilon$.  
}
\end{figure}
The charge configuration in the two dots, $(n_L,n_R)$, is controlled by the gate 
voltages $V_L$ and $V_R$. 
In particular, by controlling the dimensionless parameter $\epsilon \propto V_R -V_L$, 
the charge configuration is continuously tuned between $(0,2)$ and $(1,1)$. 
When the two electrons are in the same dot  $(0,2)$, the ground state is a spin singlet, 
$\ket{S(0,2)}$; the highly energetic excited triplet states are decoupled.  
For  $(1,1)$  the degeneracy of the triplet states is removed by a magnetic field, 
$\textbf{B}=B \hat{\textbf{z}}$, applied perpendicularly to the sample's plane; 
$\ket{S(1,1)}$ and $\ket{T_0(1,1)}$ are degenerate. 
The charging energy and the spin preserving inter-dot 
tunneling (controlled by the gate voltage $V_T$)
are described by $H_0=\Delta_s(0)[\epsilon (\ket{T_0(1,1)}\bra{T_0(1,1)}+\ket{S(1,1)}\bra{S(1,1)}-\ket{S(0,2)}\bra{S(0,2)})+1/2 (\ket{S(0,2)}\bra{S(1,1)}+\trm{H.c.})]$, where  $\Delta_s(0)/2$  is the tunneling amplitude. 
The singlet ground state,  $\ket{S_g(\epsilon)}$, 
and the excited state,  $\ket{S_e(\epsilon)}$, together with $\ket{T_0(1,1)}$,
 diagonalize the Hamiltonian (cf. Fig. 1(b))
\bea
H_0= & & \Delta_s(\epsilon)/2 \left( \ket{S_e(\epsilon)}\bra{S_e(\epsilon)} - 
\ket{S_g(\epsilon)}\bra{S_g(\epsilon)}\right) \nonumber \\ 
& & +\Delta_s(0)\epsilon \ket{T_0(1,1)}\bra{T_0(1,1)} \, .
\label{due}
\eea
The energy gap,  $J(\epsilon)=\Delta_s(\epsilon)/2+\epsilon \Delta_s(0)$, between  $\ket{S_g(\epsilon)}$ and $\ket{T_0(1,1)}$ 
is vanishingly small  at $\epsilon \lesssim \epsilon_B$  (cf. Fig. 1(b)). 
The hyperfine interaction between electrons and the nuclear
 spin~\cite{nuclei},
 facilitates transitions between these states. 
For our purpose, the effect of the nuclear spins on the electrons is 
described by classical magnetic fields, ${{\bf B}_N}_L$,  ${{\bf B}_N}_R$,
 resulting in the Hamiltonian  $H_N=g \mu_B ({{\bf B}_N}_R-{{\bf B}_N}_L) \cdot \hat{{\bf z}} \ket{T_0(1,1)}\bra{S(1,1)}+ \trm{H.c.}$.

Dealing with the spin degree of freedom enables us to achieve long dephasing times,
 which allows for weak (continuous) measurements~\cite{Jordan:2007aa}. 
 By contrast, the detectors (D1 and D2),
which are two quantum point contacts (QPCs) located near the dots, are charge sensors ~\cite{sensing} suitable for continuous measurements~\cite{Korotkov:2001aa}.
D2 is effectively used to perform a strong measurement of the spin, following a spin-to-charge
conversion, i.e. a mapping of different spin states at $\epsilon = \epsilon_B$ 
into different charge states at $\epsilon=\epsilon_A$, 
employing an adiabatic (as compared to the tunneling Hamiltonian)
variation of $\epsilon$ (cf. Fig. 1). 
D1 is used as a weak detector. It is sensitive to the difference between two spin states; the latter 
correspond to two different charge configurations. This is indeed the case for $\epsilon \neq \epsilon_B$.
The interaction between the double QD and the QPC is modeled as 
$H_{\trm{int}}=(H_{(0,2)}-H_{(0,1)})\otimes\ket{S(0,2)}\bra{S(0,2)} +H_{(1,1)}\otimes \id$. 
For the $(1,1)$  (or  $(0,2)$) charge configuration, the electrons in the QPC are 
described by the Hamiltonian  $H_{(1,1)}$ (or  $H_{(0,2)}$). 
Assuming the excited singlet state is not populated (which is the case for $k_B T$, 
$eV$, $\hbar/\tau \ll \Delta_s(\epsilon)$, cf. Fig. 1(b)), the interaction Hamiltonian 
can be written as $H_{\trm{int}} \approx H_{(1,1)}+ J(\epsilon)/\Delta_s(\epsilon)
(H_{(0,2)}-H_{(1,1)})\otimes(\id-\hat{S}^2/2)$, where the measured observable,  
$ \hat{A} \equiv \id-\hat{S}^2/2=\ket{S_g(\epsilon)}\bra{S_g(\epsilon)}$, is the singlet component 
of the spin state. 
 $H_{(1,1)}$  describes scattering of the electrons in the QPC with 
transmission (reflection) coefficient $t_0$  ($r_0$): any incoming electron 
in the QPC,  $\ket{\trm{in}}$, evolves to  $\ket{\phi}=t_0\ket{t}+r_0 \ket{r}$, 
where $\ket{t}$ and $\ket{r}$  are the reflected and transmitted states for the electron.
 If the system is in the  $\ket{S_g(\epsilon)}$ state, the electron in the QPC 
evolves according to $\ket{\trm{in}} \ra \ket{\phi'}=(t_0+\delta t(\epsilon))\ket{t}+(r_0+\delta r(\epsilon))\ket{r}=\ket{\phi}+\ket{\Delta \phi}$. $\delta t$, $\delta r$ can be tuned to be arbitrarily small in  $J(\epsilon)/\Delta_s(\epsilon)$.

{\it The protocol.}--- The protocol consists of a weak measurement with post-selection realized by a sequence of voltage pulses as described in Fig. 2(a). 
\begin{figure}[h!]
\label{fig_2}
\begin{center}
\includegraphics[width=70mm]{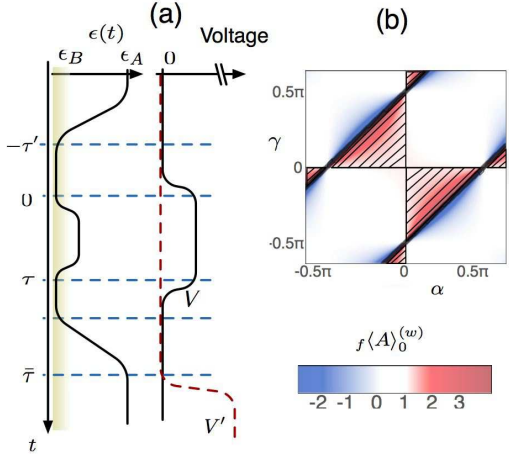}
\end{center}
\caption{(Color online) (a) A protocol to measure weak values of two electron spin: shown are
  $V$, $V'$,  (voltage bias across D1 and D2 respectively) 
and $\epsilon$.
(b) The weak value,  ${}_f\Avg{\hat{A}}_0^{(W)} =[1-{}_f\Avg{\hat{S}^2}_0^{(W)}]/2$, as a function of the parameters $\alpha$ and $\gamma$  for  $\beta=\pi$. 
The dark region defines the range of parameters for which a positive post-selection is obtained with probability  $P_{(0,2)}<0.5\% $.  The shadowed region (parallel lines) corresponds to the values of the parameters for which  ${}_f\Avg{\hat{S}^2}_0^{(W)}<0$. 
}
\end{figure}
The evolution of the system in the absence of the detector has already been realized in 
experiment~\cite{Petta:2005aa}. 
Initially, at  $\epsilon=\epsilon_A$, the system is in the ground state,  $\ket{S(0,2)}$. 
By a fast adiabatic variation (cf. Fig. 1) it is evolved into 
 $\ket{S(1,1)}$ ($\epsilon=\epsilon_B$ at time $t=-\tau'$). 
This state evolves under the influence of the nuclear interaction until time  $t=0$, 
thus preselecting  $\ket{\chi_0}=\cos \alpha \ket{S_g}-i \sin \alpha \ket{T_0}$ with
  $\alpha=g \mu_B ({{\bf B}_N}_R-{{\bf B}_N}_L) \cdot \hat{{\bf z}} \tau'$. 
During the measurement pulse the free evolution of the system is 
 $\mathcal{U}(\tau,0)= \ket{S_g}\bra{S_g}+ \exp(-i \beta) \ket{T_0}\bra{T_0}$,
 with $\beta=J(\epsilon)\tau$. 
The evolution during the time interval $(\tau, \tau+\tau'')$ 
 is governed by the nuclear interaction and that from $\tau+\tau''$ 
 to $\bar{\tau}$  is a fast adiabatic variation. 
The evolution  $\mathcal{U}(\bar{\tau},\tau)$ defines the effective 
post-selected state  $ |\chi_f' \left. \right\rangle=\mathcal{U}^{-1}(\bar{\tau},\tau)
\ket{S(0,2)}=\cos \gamma \ket{S_g}+ i \sin \gamma \ket{T_0}$,
 where  $\gamma=g \mu_B ({\bf B}_{NR}-{\bf B}_{NL}) \cdot \hat{{\bf z}} \tau''$ (cf. Fig.~2(b)).
The weak value of $\mathcal{U}(\tau,0)\hat{A}$ is then 
${}_f\Avg{\hat{A}}_0^{(W)}=\cos \gamma \cos \alpha /(\cos \gamma \cos \alpha - 
e^{-i \beta} \sin \gamma \sin \alpha)$. 
 By tuning the duration of the pulses, one can obtain a real WV 
(e.g. for  $\beta=\pi$), which is arbitrarily large (e.g. for  $\gamma-\alpha \ra \pi/2$) --cf. Fig. 2(b).

The interaction with the detector in this protocol is depicted  by a simple model, 
in which the electrons in the double dot interact with a single electron in the QPC. 
Once the state  $\ket{\chi_0}$ is prepared at time  $t=0$, the system interacts 
with the QPC, $t \in [0, \tau]$, creating an entangled state at time  $t=\tau$, 
$\ket{\psi}=\mathcal{U}(\tau,0)\ket{\chi_0}\ket{\phi}+ \hat{A}\mathcal{U}(\tau,0)
\ket{\chi_0}\ket{\Delta\phi}$, where  $\mathcal{U}(\tau,0)$ defines the time 
evolution of the system from $t=0$  to  $t=\tau$. 
Applying the operator  $\Pi_t \equiv \ket{t}\bra{t}$ one can now detect whether the electron in the QPC has ($n=1$) or has not ($n=0$) been transmitted. The respective 
probabilities are $P(n=1)=\Avg{\psi | \Pi_t | \psi}$,
$P(n=0)=1-P(n=1)$. In either case, the corresponding spin
state of the system is given by $\ket{\bar{\chi}_n}=[\Pi_t^n+(1-\Pi_t)^{(1-n)}]\ket{\psi} / \sqrt{\Avg{\psi | \Pi_t | \psi}}$.
We employ QPC D2 to detect the charge configuration in the double 
dot (post selection) at a later time $\bar{\tau}$ (cf. Fig. 2(a)), but use 
time  $t=\tau$ to express the post-selected state, 
 $|\chi_f'\rangle=\mathcal{U}^{-1}(\bar{\tau},\tau)\ket{\chi_f}$, 
in terms of the time evolved state, $\ket{\chi_f}$  at time $t=\bar{\tau}$. 
The signal of QPC D1 is kept, provided D2 measures the 
$(0,2)$ charge configuration (with probability 
$P((0,2)|n)=|\Avg{\bar{\chi}_n|\chi_f}|^2$). 
This corresponds to averaging the reading of the 
first QPC conditional to the positive outcome ($(0,2)$ 
charge configuration) of the second measurement, 
${}_f\Avg{n}_0=\sum_{n=0,1}n P(n)P((0,2)|n)/(\sum_n P(n)P((0,2)|n))$. 
If  $\mbox{\rm Re} \{ {}_f\Avg{\hat{A}}_0^{(W)} t_0^* \delta t\} \ll 1$, 
the average number of transmitted electrons is 
\bea
& & {}_f\Avg{n}_0= \left| t_0 \right|^2 +2 \mbox{\rm Re} \left\{ {}_f\Avg{\hat{A}}_0^{(W)} t_0^* \delta t \right\}  \, , \nonumber \\
& &  {}_f\Avg{\hat{A}}_0^{(W)}=\Avg{\chi_f'|\mathcal{U}(\tau,0)\hat{A}|\chi_0}/\Avg{\chi_f'|\mathcal{U}(\tau,0) |\chi_0} \, ,
\label{tre}
\eea
defining the weak value ${}_f\Avg{\hat{A}}_0^{(W)}$. 
Indeed the inferred weak measurement operator,  
$\mathcal{U}(\tau,0)\hat{A}$, and the strong post-selection operator,  
$\mathcal{U}^{-1}(\bar{\tau},\tau) \hat{A} \mathcal{U}(\bar{\tau},\tau)$, 
both expressed at time  $\tau$, do not commute with each other, as required to obtain nonstandard weak values.  
This measurement may capture the real and the imaginary part of the WV; 
one can reconstruct the complex WV provided the phase of  
$t_0^* \delta t$ is tunable in a controlled way. 
In particular this is possible if one embeds the QPC in an interferometry device. 
Note that in the absence of post-selection, \eq{tre} is replaced by
 $n=|t_0|^2 + 2 \Avg{\hat{A}}\mbox{\rm Re} \{ t_0^* \delta t\}$.

The result of this simple model captures the physics of weak values. 
Indeed, during the measurement time,  $\tau$, the number of electrons attempting 
to pass through the QPC is  $N=2eV\tau/(2\pi \hbar)$. 
In this case the probability that  $n$ electrons out of  $N$ will pass 
through the QPC is  $P(n,N)=|\Avg{S_g(\epsilon)| \chi_0}|^2 P_{t_0+\delta t}(n,N)+
|\Avg{T_0(1,1)|\chi_0}|^2 P_{t_0}(n,N)$, with  $P_{x}(n,N)=N!/[n! (N-n)!] |x|^{2n} 
(1-|x|^2)^{(N-n)}$.
 If  $|1+t_0^* \delta t +r_0^* \delta r|^N \ll 1$, the two distribution functions, 
$P_{t_0}$,  $P_{t_0+\delta t}$, are strongly overlapping and the average current in the 
QPC,  ${}_f\Avg{I}_0=e\hspace{0.1cm}{}_f\Avg{n}_0/\tau$, will measure, to leading 
order in $t_0^* \delta t$, the WV  ${}_f\Avg{I}_0=I_0+(2 e^2V/h)2
 \mbox{\rm Re} \{ {}_f\Avg{\hat{A}}_0^{(W)} t_0^* \delta t \}$. Here  $I_0=2 e^2V|t_0|^2/h$
 is the current for the   $(1,1)$ charge configuration. 
Note that this result essentially coincides with that of the simplified
 picture outlined above. In the opposite limit, 
 $|1+t_0^* \delta t +r_0^* \delta r|^N \gg 1$, the  overlap between  
$P_{t_0}$ and  $P_{t_0+\delta t}$ is vanishing, in which case the measurement 
is strong: the outcome of each single measurement is either
 $2e^2V|t_0|^2/h$  or $2 e^2V|t_0+\delta t|^2/h$. Note that the parameter
 controlling the crossover from weak to strong measurement is the same 
controlling the decoherence of the double dot state due to the 
measurement~\cite{Averin:2005aa}.

{\it Weak values protected from nuclear field induced decoherence.}--- Weak values are 
sensitive to decoherence effects. 
The latter arise not only from the measuring device itself. In the protocol discussed
 above decoherence is dominated by fluctuations of the nuclear 
spins~\cite{Johnson:2005aa, Petta:2005aa}. 
While new emerging experimental techniques carry the promise of an 
increased level of coherent control~\cite{recent}, a 
protocol realizable in actual experiments which is insensitive to nuclear 
spin fluctuations is depicted in Fig. 3. 
\begin{figure}[ht!]
\label{fig_3}
\begin{center}
\includegraphics[width=70mm]{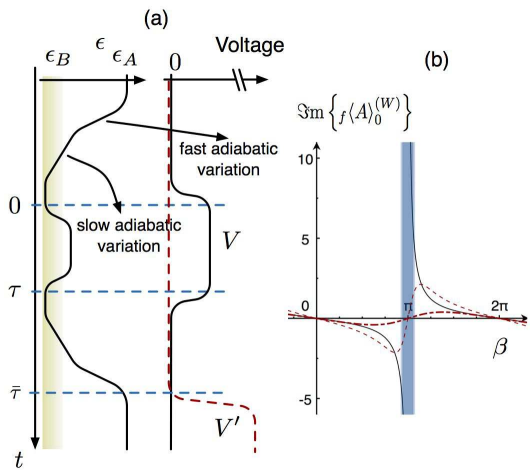}
\end{center}
\caption{(Color online) (a) An alternative protocol to measure ${}_f\Avg{\hat{A}}_0^{(W)}$: shown are $\epsilon$,  $V$ and $V'$ (notation as in Fig. 2). 
(b) Imaginary part of the weak value of the spin as a function of  $\beta$ for $\Gamma \tau=0$ (full line), $\Gamma \tau=0.1$ (dashed curve) and $\Gamma \tau =1$  (dash-dotted curve).
 The values of parameters for which the corresponding weak values of spin are obtained with probability  $P_{(0,2)}<1\%$ are shadowed. 
}
\end{figure}
Here, though the freedom in 
defining the pre- and post-selected state is reduced: one is restricted to
 move only within the equatorial plane of the Bloch sphere.
The protocol consists in starting with the system in the ground state,  
$\ket{S(0,2)}$ (at  $\epsilon=\epsilon_A$), a slow adiabatic variation 
(cf. Fig. 1) allows the system to evolve to $\ket{\uparrow \downarrow}$  
at time $t=0$  ($\epsilon=\epsilon_B$). 
The time evolution during the weak measurement pulse is  
$\mathcal{U}(\tau,0)=\ket{S_g}\bra{S_g}+\exp(-i \beta) \ket{T_0}\bra{T_0}$,
 with $\beta=J(\epsilon) \tau$. 
The evolution from $\tau$  to $\bar{\tau}$ is ``slow adiabatic''. 
The effective post-selected state at time  $\tau$ is  $ |\chi_f'\left. 
\right\rangle=\mathcal{U}^{-1}(\bar{\tau},\tau) \ket{\chi_f}$. 
The weak value of $\mathcal{U}(\tau,0)\hat{A}$ is then 
${}_f\Avg{\hat{A}}_0^{(W)}=1/2[1-i \tan(\beta/2)]$. The imaginary part of the
 weak value can then be arbitrary large. 
 Decoherence due to fluctuations of the electromagnetic field  
is however unavoidable~\cite{mio}.
 A general scheme to account for the effects of decoherence requires 
the use of a density matrix, $\rho(\tau)$.  
In the present protocol decoherence mainly comes from charge 
fluctuations (fluctuations of  $\epsilon$), which 
commute with the measured operator $\hat{A}$, yielding 
${}_f\Avg{\hat{A}}_0^{(W)}=\Avg{\chi_f'|\rho(\tau)\hat{A}|\chi_f'}/
\Avg{\chi_f'| \rho(\tau)|\chi_f'}$. 
Here the density matrix is $\rho(\tau)=1/2(\ket{S_g}\bra{S_g}+\ket{T_0}\bra{T_0}
+ e^{-i(\beta-i\Gamma \tau)}\ket{S_g}\bra{T_0}+e^{+i(\beta+i\Gamma \tau)}\ket{T_0}\bra{S_g})$, 
where $\Gamma$  is defined by  $J(\epsilon) \ra J(\epsilon)+\xi(t)$, 
with  $\Avg{\xi(t)\xi(t')}=2 \hbar\Gamma \delta (t-t')$ --cf. Fig. 3. 
In general the strong incoherent 
limit does not reduce to the standard expectation value of the spin.
The presence of coherent oscillations within this protocol~\cite{Petta:2005aa}, 
anyway, underlines the possibility to realize this procedure experimentally. 
The present protocol employs building blocks which have already been tested experimentally (QPCs as charge sensors, spin manipulation in the double dot). Most promising is the experiment setup of Ref.~\cite{Petta:2005aa} . The main experimental future challenge here would be a single shot readout, i.e. a quantum mechanical measurement of the state of the dot (without averaging over many repetitions as in ref. [13]). In this context we note that charge sensing by a QPC operated with fast pulses has been recently demonstrated~\cite{Reilly:2007aa}.

{\it Conclusions.}--- The protocol outlined above will facilitate the measurement 
of non-standard (weak) values of spin. 
The procedure is generalized to include the effect of non-pure states. 
Further directions may include a systematic study of various mechanisms for 
decoherence within the weak value scheme and the measurement of two 
interacting spins (pair of double quantum dots), leading to cross-correlations 
of weak values.

We are grateful to Y. Aharonov for introducing us to the subject of weak values. 
We acknowledge useful discussions with Y. Aharonov, L. Vaidman and A. Yacoby on 
theoretical and experimental aspects of the problem. This work was supported in 
part by the U.S.-Israel BSF, the DFG project SPP 1285, the Transnational 
Access Program RITA-CT-2003-506095 and the Minerva Foundation.

{\it Note added.}--- Upon submission of this manuscript we have noted the paper by Williams 
and Jordan~\cite{Williams:2008aa} which too discusses weak values in the context 
of solid state devices.


\begin{thebibliography}{25}
\expandafter\ifx\csname natexlab\endcsname\relax\def\natexlab#1{#1}\fi
\expandafter\ifx\csname bibnamefont\endcsname\relax
  \def\bibnamefont#1{#1}\fi
\expandafter\ifx\csname bibfnamefont\endcsname\relax
  \def\bibfnamefont#1{#1}\fi
\expandafter\ifx\csname citenamefont\endcsname\relax
  \def\citenamefont#1{#1}\fi
\expandafter\ifx\csname url\endcsname\relax
  \def\url#1{\texttt{#1}}\fi
\expandafter\ifx\csname urlprefix\endcsname\relax\def\urlprefix{URL }\fi
\providecommand{\bibinfo}[2]{#2}
\providecommand{\eprint}[2][]{\url{#2}}

\bibitem[{\citenamefont{von Neuman}(1932)}]{Neumann1932}
\bibinfo{author}{\bibfnamefont{J.}~\bibnamefont{von Neuman}},
  \emph{\bibinfo{title}{Mathematische Grusndlagen der Quantemechanik}}
  (\bibinfo{publisher}{Springler-Verlag, Berlin}, \bibinfo{year}{1932}).

\bibitem[{\citenamefont{Aharonov et~al.}(1964)\citenamefont{Aharonov, Bergmann,
  and Lebowitz}}]{Aharonov:1964aa}
\bibinfo{author}{\bibfnamefont{Y.}~\bibnamefont{Aharonov}},
  \bibinfo{author}{\bibfnamefont{P.}~\bibnamefont{Bergmann}}, \bibnamefont{and}
  \bibinfo{author}{\bibfnamefont{J.}~\bibnamefont{Lebowitz}},
  \bibinfo{journal}{Phys. Rev.} \textbf{\bibinfo{volume}{134}},
  \bibinfo{pages}{B1410} (\bibinfo{year}{1964}).

\bibitem[{\citenamefont{Aharonov et~al.}(1988)\citenamefont{Aharonov, Albert,
  and Vaidman}}]{Aharonov:1988aa}
\bibinfo{author}{\bibfnamefont{Y.}~\bibnamefont{Aharonov}},
  \bibinfo{author}{\bibfnamefont{D. Z.}~\bibnamefont{Albert}}, \bibnamefont{and}
  \bibinfo{author}{\bibfnamefont{L.}~\bibnamefont{Vaidman}},
  \bibinfo{journal}{Phys. Rev. Lett.} \textbf{\bibinfo{volume}{60}},
  \bibinfo{pages}{1351} (\bibinfo{year}{1988}).

\bibitem[{\citenamefont{Aharonov and Vaidman}(1990)}]{Aharonov:1990aa}
\bibinfo{author}{\bibfnamefont{Y.}~\bibnamefont{Aharonov}} \bibnamefont{and}
  \bibinfo{author}{\bibfnamefont{L.}~\bibnamefont{Vaidman}},
  \bibinfo{journal}{Phys. Rev. A} \textbf{\bibinfo{volume}{41}},
  \bibinfo{pages}{11} (\bibinfo{year}{1990}).

\bibitem[{200(2002)}]{2002proc.}
\bibinfo{author}{\bibfnamefont{Y.}~\bibnamefont{Aharonov}}, \bibnamefont{and}
\bibinfo{author}{\bibfnamefont{L.}~\bibnamefont{Vaidman}}, \bibnamefont{in}
\emph{\bibinfo{title}{Time in Quantum Mechanics}}, J.G. Muga et al. eds., (Springer)  369-412,(\bibinfo{year}{2002}).

\bibitem{non_commuting}
\bibinfo{author}{\bibfnamefont{A.~N.}~\bibnamefont{Jordan}} \bibnamefont{and}
  \bibinfo{author}{\bibfnamefont{M.}~\bibnamefont{Buttiker}},
  \bibinfo{journal}{Phys. Rev. Lett.} \textbf{\bibinfo{volume}{95}},
  \bibinfo{pages}{220401} (\bibinfo{year}{2005}); 
\bibinfo{author}{\bibfnamefont{W.}~\bibnamefont{Hongduo}} \bibnamefont{and}
  \bibinfo{author}{\bibfnamefont{Y. V.}~\bibnamefont{Nazarov}},
  \bibinfo{journal}{cond-mat/0703344}  (\bibinfo{year}{2007}).

\bibitem[{\citenamefont{Neder et~al.}(2007)\citenamefont{Neder, Heiblum,
  Mahalu, and Umansky}}]{Neder:2007aa}
\bibinfo{author}{\bibfnamefont{I.}~\bibnamefont{Neder}},
  \bibinfo{author}{\bibfnamefont{M.}~\bibnamefont{Heiblum}},
  \bibinfo{author}{\bibfnamefont{D.}~\bibnamefont{Mahalu}}, \bibnamefont{and}
  \bibinfo{author}{\bibfnamefont{V.}~\bibnamefont{Umansky}},
  \bibinfo{journal}{Phys. Rev. Lett.} \textbf{\bibinfo{volume}{98}},
  \bibinfo{pages}{036803} (\bibinfo{year}{2007}).

\bibitem{correlations}
\bibinfo{author}{\bibfnamefont{A.} \bibnamefont{Di Lorenzo}} \bibnamefont{and}
  \bibinfo{author}{\bibfnamefont{Y.~V.} \bibnamefont{Nazarov}},
  \bibinfo{journal}{Phys. Rev. Lett.} \textbf{\bibinfo{volume}{93}},
  \bibinfo{pages}{046601} (\bibinfo{year}{2004});
\bibinfo{author}{\bibfnamefont{E.}~\bibnamefont{Sukhorukov}},
  \bibinfo{author}{\bibfnamefont{A.}~\bibnamefont{Jordan}},
  \bibinfo{author}{\bibfnamefont{S.}~\bibnamefont{Gustavsson}},
  \bibinfo{author}{\bibfnamefont{R.}~\bibnamefont{Leturcq}},
  \bibinfo{author}{\bibfnamefont{T.}~\bibnamefont{Ihn}}, \bibnamefont{and}
  \bibinfo{author}{\bibfnamefont{K.}~\bibnamefont{Ensslin}},
  \bibinfo{journal}{Nat Phys} \textbf{\bibinfo{volume}{3}},
  \bibinfo{pages}{243} (\bibinfo{year}{2007}),
  \bibinfo{note}{10.1038/nphys564}.
  
  
\bibitem{optics}
\bibinfo{author}{\bibfnamefont{N.~W.~M.} \bibnamefont{Ritchie}},
  \bibinfo{author}{\bibfnamefont{J.~G.} \bibnamefont{Story}}, \bibnamefont{and}
  \bibinfo{author}{\bibfnamefont{R.~G.} \bibnamefont{Hulet}}, 
  \bibinfo{journal}{Phys. Rev. Lett.} \textbf{\bibinfo{volume}{66}},
  \bibinfo{pages}{1107} (\bibinfo{year}{1991});
\bibinfo{author}{\bibfnamefont{A. M.}~\bibnamefont{Steinberg}},
  \bibinfo{journal}{{\it ibid.}} \textbf{\bibinfo{volume}{74}},
  \bibinfo{pages}{2405} (\bibinfo{year}{1995});
\bibinfo{author}{\bibfnamefont{G.~J.} \bibnamefont{Pryde}},
  \bibinfo{author}{\bibfnamefont{J.~L.} \bibnamefont{O'Brien}},
  \bibinfo{author}{\bibfnamefont{A.~G.} \bibnamefont{White}},
  \bibinfo{author}{\bibfnamefont{T.~C.} \bibnamefont{Ralph}}, \bibnamefont{and}
  \bibinfo{author}{\bibfnamefont{H.~M.} \bibnamefont{Wiseman}},
  \bibinfo{journal}{{\it ibid.}} \textbf{\bibinfo{volume}{94}},
  \bibinfo{pages}{220405} (\bibinfo{year}{2005}).

\bibitem[{\citenamefont{Petta et~al.}(2005)\citenamefont{Petta, Johnson,
  Taylor, Laird, Yacoby, Lukin, Marcus, Hanson, and Gossard}}]{Petta:2005aa}
\bibinfo{author}{\bibfnamefont{J.~R.} \bibnamefont{Petta}},
  \bibinfo{author}{\bibfnamefont{A.~C.} \bibnamefont{Johnson}},
  \bibinfo{author}{\bibfnamefont{J.~M.} \bibnamefont{Taylor}},
  \bibinfo{author}{\bibfnamefont{E.~A.} \bibnamefont{Laird}},
  \bibinfo{author}{\bibfnamefont{A.}~\bibnamefont{Yacoby}},
  \bibinfo{author}{\bibfnamefont{M.~D.} \bibnamefont{Lukin}},
  \bibinfo{author}{\bibfnamefont{C.~M.} \bibnamefont{Marcus}},
  \bibinfo{author}{\bibfnamefont{M.~P.} \bibnamefont{Hanson}},
  \bibnamefont{and} \bibinfo{author}{\bibfnamefont{A.~C.}
  \bibnamefont{Gossard}}, \bibinfo{journal}{Science}
  \textbf{\bibinfo{volume}{309}}, \bibinfo{pages}{2180} (\bibinfo{year}{2005}).

\bibitem{nuclei}
\bibinfo{author}{\bibfnamefont{S.~I.}~\bibnamefont{Erlingsson}},
  \bibinfo{author}{\bibfnamefont{Y.~V.}~\bibnamefont{Nazarov}}, \bibnamefont{and}
  \bibinfo{author}{\bibfnamefont{V.~I.}~\bibnamefont{Fal'ko}},
  \bibinfo{journal}{Phys. Rev. B} \textbf{\bibinfo{volume}{64}},
  \bibinfo{pages}{195306} (\bibinfo{year}{2001});
\bibinfo{author}{\bibfnamefont{A.~V.} \bibnamefont{Khaetskii}},
  \bibinfo{author}{\bibfnamefont{D.}~\bibnamefont{Loss}}, \bibnamefont{and}
  \bibinfo{author}{\bibfnamefont{L.}~\bibnamefont{Glazman}},
  \bibinfo{journal}{Phys. Rev. Lett.} \textbf{\bibinfo{volume}{88}},
  \bibinfo{pages}{186802} (\bibinfo{year}{2002});
\bibinfo{author}{\bibfnamefont{I.~A.}~\bibnamefont{Merkulov}},
  \bibinfo{author}{\bibfnamefont{A.~L.}~\bibnamefont{Efros}}, \bibnamefont{and}
  \bibinfo{author}{\bibfnamefont{M.}~\bibnamefont{Rosen}},
  \bibinfo{journal}{Phys. Rev. B} \textbf{\bibinfo{volume}{65}}
  \bibinfo{pages}{205309} (\bibinfo{year}{2002}).

\bibitem[{\citenamefont{Jordan et~al.}(2007)\citenamefont{Jordan, Trauzettel,
  and Burkard}}]{Jordan:2007aa}
\bibinfo{author}{\bibfnamefont{A.}~\bibnamefont{Jordan}},
  \bibinfo{author}{\bibfnamefont{B.}~\bibnamefont{Trauzettel}},
  \bibnamefont{and} \bibinfo{author}{\bibfnamefont{G.}~\bibnamefont{Burkard}},
  \bibinfo{journal}{arXiv:0706.0180}  (\bibinfo{year}{2007}).

\bibitem{sensing}
\bibinfo{author}{\bibfnamefont{M.}~\bibnamefont{Field}},
  \bibinfo{author}{\bibfnamefont{C.~G.}~\bibnamefont{Smith}},
  \bibinfo{author}{\bibfnamefont{M.}~\bibnamefont{Pepper}},
  \bibinfo{author}{\bibfnamefont{D.~A.}~\bibnamefont{Ritchie}},
  \bibinfo{author}{\bibfnamefont{J.~E.~F.}~\bibnamefont{Frost}},
  \bibinfo{author}{\bibfnamefont{G.~A.~C.}~\bibnamefont{Jones}}, \bibnamefont{and}
  \bibinfo{author}{\bibfnamefont{D.~G.}~\bibnamefont{Hasko}},
  \bibinfo{journal}{Phys. Rev. Lett.} \textbf{\bibinfo{volume}{70}},
  \bibinfo{pages}{1311} (\bibinfo{year}{1993});
\bibinfo{author}{\bibfnamefont{L.}~\bibnamefont{DiCarlo}},
  \bibinfo{author}{\bibfnamefont{H.~J.} \bibnamefont{Lynch}},
  \bibinfo{author}{\bibfnamefont{A.~C.} \bibnamefont{Johnson}},
  \bibinfo{author}{\bibfnamefont{L.~I.} \bibnamefont{Childress}},
  \bibinfo{author}{\bibfnamefont{K.}~\bibnamefont{Crockett}},
  \bibinfo{author}{\bibfnamefont{C.~M.} \bibnamefont{Marcus}},
  \bibinfo{author}{\bibfnamefont{M.~P.} \bibnamefont{Hanson}},
  \bibnamefont{and} \bibinfo{author}{\bibfnamefont{A.~C.}
  \bibnamefont{Gossard}}, \bibinfo{journal}{{\it ibid.}}
  \textbf{\bibinfo{volume}{92}}, \bibinfo{pages}{226801}
  (\bibinfo{year}{2004}).

\bibitem{Korotkov:2001aa}
\bibinfo{author}{\bibfnamefont{A.~N.} \bibnamefont{Korotkov}} \bibnamefont{and}
  \bibinfo{author}{\bibfnamefont{D.~V.} \bibnamefont{Averin}},
  \bibinfo{journal}{Phys. Rev. B} \textbf{\bibinfo{volume}{64}},
  \bibinfo{pages}{165310} (\bibinfo{year}{2001}).


\bibitem[{\citenamefont{Averin and Sukhorukov}(2005)}]{Averin:2005aa}
\bibinfo{author}{\bibfnamefont{D.~V.} \bibnamefont{Averin}} \bibnamefont{and}
  \bibinfo{author}{\bibfnamefont{E.~V.} \bibnamefont{Sukhorukov}},
  \bibinfo{journal}{Phys. Rev. Lett.} \textbf{\bibinfo{volume}{95}},
  \bibinfo{pages}{126803} (\bibinfo{year}{2005}).
  
\bibitem[{\citenamefont{Johnson et~al.}(2005)\citenamefont{Johnson, Petta,
  Taylor, Yacoby, Lukin, Marcus, Hanson, and Gossard}}]{Johnson:2005aa}
\bibinfo{author}{\bibfnamefont{A.~C.} \bibnamefont{Johnson}},
  \bibinfo{author}{\bibfnamefont{J.~R.} \bibnamefont{Petta}},
  \bibinfo{author}{\bibfnamefont{J.~M.} \bibnamefont{Taylor}},
  \bibinfo{author}{\bibfnamefont{A.}~\bibnamefont{Yacoby}},
  \bibinfo{author}{\bibfnamefont{M.~D.} \bibnamefont{Lukin}},
  \bibinfo{author}{\bibfnamefont{C.~M.} \bibnamefont{Marcus}},
  \bibinfo{author}{\bibfnamefont{M.~P.} \bibnamefont{Hanson}},
  \bibnamefont{and} \bibinfo{author}{\bibfnamefont{A.~C.}
  \bibnamefont{Gossard}}, \bibinfo{journal}{Nature}
  \textbf{\bibinfo{volume}{435}}, \bibinfo{pages}{925} (\bibinfo{year}{2005}).

\bibitem{recent}
\bibinfo{author}{\bibfnamefont{F.~H.~L.} \bibnamefont{Koppens}},
  \bibinfo{author}{\bibfnamefont{C.}~\bibnamefont{Buizert}},
  \bibinfo{author}{\bibfnamefont{K.~J.} \bibnamefont{Tielrooij}},
  \bibinfo{author}{\bibfnamefont{I.~T.} \bibnamefont{Vink}},
  \bibinfo{author}{\bibfnamefont{K.~C.} \bibnamefont{Nowack}},
  \bibinfo{author}{\bibfnamefont{T.}~\bibnamefont{Meunier}},
  \bibinfo{author}{\bibfnamefont{L.~P.} \bibnamefont{Kouwenhoven}},
  \bibnamefont{and} \bibinfo{author}{\bibfnamefont{L.~M.~K.}
  \bibnamefont{Vandersypen}}, \bibinfo{journal}{Nature}
  \textbf{\bibinfo{volume}{442}}, \bibinfo{pages}{766} (\bibinfo{year}{2006});
\bibinfo{author}{\bibfnamefont{E.}~\bibnamefont{Laird}},
  \bibinfo{author}{\bibfnamefont{C.}~\bibnamefont{Barthel}},
  \bibinfo{author}{\bibfnamefont{E.}~\bibnamefont{Rashba}},
  \bibinfo{author}{\bibfnamefont{C.}~\bibnamefont{Marcus}},
  \bibinfo{author}{\bibfnamefont{M.}~\bibnamefont{Hanson}}, \bibnamefont{and}
  \bibinfo{author}{\bibfnamefont{A.}~\bibnamefont{Gossard}},
  \bibinfo{journal}{cond-mat/0707.0557}  (\bibinfo{year}{2007}).

\bibitem{mio}
\bibinfo{author}{\bibfnamefont{G.}~\bibnamefont{Burkard}},
  \bibinfo{author}{\bibfnamefont{D.}~\bibnamefont{Loss}}, \bibnamefont{and}
  \bibinfo{author}{\bibfnamefont{D.~P.}~\bibnamefont{DiVincenzo}},
  \bibinfo{journal}{Phys. Rev. B} \textbf{\bibinfo{volume}{59}},
  \bibinfo{pages}{2070} (\bibinfo{year}{1999});
\bibinfo{author}{\bibfnamefont{A.}~\bibnamefont{Romito}} \bibnamefont{and}
  \bibinfo{author}{\bibfnamefont{Y.}~\bibnamefont{Gefen}},
  \bibinfo{journal}{{\it ibid.}} \textbf{\bibinfo{volume}{76}}, 
  \bibinfo{pages}{195318} (\bibinfo{year}{2007}).

\bibitem[{\citenamefont{Reilly, Marcus, Hanson, Gossard}(2007)}]{Reilly:2007aa}
\bibinfo{author}{\bibfnamefont{D.~J.}~\bibnamefont{Reilly}},
\bibinfo{author}{\bibfnamefont{C.~M.}~\bibnamefont{Marcus}},
\bibinfo{author}{\bibfnamefont{M.~P.}~\bibnamefont{Hanson}}, \bibnamefont{and}
  \bibinfo{author}{\bibfnamefont{A.~C.}~\bibnamefont{Gossard}},
  \bibinfo{journal}{cond-mat/0707.2946}  (\bibinfo{year}{2007}).

\bibitem{Williams:2008aa}
 \bibinfo{author}{\bibfnamefont{N. S.}~\bibnamefont{Williams}}, \bibnamefont{and}
  \bibinfo{author}{\bibfnamefont{A. N.}~\bibnamefont{Jordan}},
  arXiv:0707.3427 [Phys. Rev. Lett. (to be published)].


\end{thebibliography}

\end{document}